\documentclass[12pt, a4paper, dvips]{article}

\usepackage{graphicx}
\usepackage{amssymb}

\def\be{\begin{equation}} \def\bea{\begin{eqnarray}}
\def\ee{\end{equation}}\def\eea{\end{eqnarray}}
\def\ncr{\nonumber\\ }

\def\a{\alpha}
\def\b{\beta}

\def\cg{{\cal G}}

\def\cl{{\cal L}}

\def\car{{\cal R}}

\def\bo{{\raise.05ex\hbox{\large$\Box$}\:}}             
\def\cbo{{\,\raise-.15ex\Sc [\,}}                       
\def\pa{\partial}                                       
\def\TH{{\raise.2ex\hbox{$\displaystyle \bigodot$}\mskip-4.7mu \llap H \;}}
\def\face{\hbox{\normalsize$\;\;\:{\raise.9ex\hbox{\oo n}\mskip-13mu \llap
        {${\buildrel{\hbox{\frtnrm ..}}\over\smile}$}}\:$}}     
\def\Face{{\raise.2ex\hbox{$\displaystyle \bigodot$}\mskip-2.2mu \llap {$\ddot
        \smile$}}}                                      
\def\Lhat{{\bf\rlap{\kern-.09em$\hat{\phantom L}$}L}}
\def\Lcheck{{\bf\rlap{\kern-.09em$\check{\phantom L}$}L}}

\def\leftrightarrowfill{$\mathsurround=0pt \mathord\leftarrow \mkern-6mu
        \cleaders\hbox{$\mkern-2mu \mathord- \mkern-2mu$}\hfill
        \mkern-6mu \mathord\rightarrow$}
\def\dvec#1{\vbox{\ialign{##\crcr
        \leftrightarrowfill\crcr\noalign{\kern-1pt\nointerlineskip}
        $\hfil\displaystyle{#1}\hfil$\crcr}}}           
\def\ddt#1{{\buildrel {\hbox{\LARGE .\kern-2pt.}} \over {#1}}}

\def\NP{Nucl. Phys. B}
\def\PLB{Phys. Lett. B}
\def\PRL{Phys. Rev. Lett.}
\def\PRD{Phys. Rev. D}

\def\MPL{Mod. Phys. Lett.}

\begin{document}

\begin{center}
{\Large\bf Dyons in Nonabelian Born--Infeld Theory}

\bigskip
{\large A. Bala\v z\footnote{E-mail: antun@phy.bg.ac.yu},
M. Buri\' c\footnote{E-mail: majab@ff.bg.ac.yu} and
V. Radovanovi\' c\footnote{E-mail: rvoja@ff.bg.ac.yu}}

\smallskip
$^1$ {\it Institute of Physics, P. O. Box 57, 11001 Belgrade, Yugoslavia}\\
$^{2, 3}$ {\it Faculty of Physics, University of Belgrade, P. O. Box 368,\\
11001 Belgrade, Yugoslavia}
\end{center}

\begin{abstract}\noindent
We analyze a nonabelian extension of Born--Infeld action for the
$SU(2)$ group. In the class of spherically symmetric solutions we
find that, besides the Gal'tsov--Kerner glueballs, only the
analytic dyons have finite energy. The presented analytic and
numerical investigation excludes the existence of pure magnetic
monopoles of 't Hooft--Polyakov type.

\vspace*{5mm}
\noindent PACS numbers: 11.10.Lm, 11.27.+d, 12.38.Lg
\end{abstract}

\section{Introduction}

Born-Infeld (BI) electrodynamics \cite{bi},  was proposed in
1934. as a theory in which the energy of  electrically charged
point particle is finite, in contrast to the Maxwell
electrodynamics. The Born-Infeld action is built similarly to the
action of relativistic point particle and it introduces
dimensional parameter, $\b$, - the ``maximal field strength". It
is usually written in one of the following forms
\bea S_{BI} =&-\, \b^2\int d^4x\left( \sqrt{-\det \left(g
_{\mu\nu}+{1\over \b }F_{\mu\nu}\right)} -
\sqrt{-\det g _{\mu\nu}}\right)\label{det}\quad\quad\quad\; \\
=&-\, \b^2\int d^4x \sqrt{-g}\left(
 \sqrt{1+{1\over 2 \b^2}F_{\mu\nu}F^{\mu\nu}-{1\over 16
\b^4}\, (F_{\mu\nu}{F^*}^{\mu\nu})^2}-1\right)\ , \label{sqrt} \eea
where  $*$ is the Hodge-dual. This action has many  interesting
properties \cite{bb}, among them duality symmetry, physical
propagation, absence of birefringence, etc.

Actions of the BI-type arise in string/M theory in two main
contexts. BI action represents the non-derivative part of the
effective open string action. As it was shown in \cite{ftt}, the
bosonic field partition function for the open string in an
external field  reduces to the BI lagrangian in the string theory
limit. On the other hand, BI action is related to D-branes. This
comes from the result that the effective action for the open
strings ending on D-branes, after the integration of string
degrees of freedom \cite{pol,gib}, is  Dirac-Born-Infeld (DBI)
action:
\be S_{DBI} = -\int d^{p+1}x \sqrt{ -\det (\eta
_{\mu\nu}+F_{\mu\nu}+\pa _\mu y^i \pa _\nu y^i )} \ , \label{DBI}\ee
where $F_{\mu\nu}$ is the field strength and
$y^i$'s are scalar fields. BI action is obtained from (\ref{DBI})
for $y^i=0$. Conversely, DBI action can be related to BI action in
higher dimensions by dimensional reduction.

The generalization of BI electrodynamics to  nonabelian gauge
theory is not unique. In the general case, if $F_{\mu\nu}$ is the
field strength of the nonabelian gauge group $\cg$ and
$F_{\mu\nu}=F_{\mu\nu}^a T_a$ ($T_a$ are the generators of $\cg$,
$[ T_a,T_b] =i f_{abc}T_c$), the "determinant" form of the action
(\ref{det}) is not equal to the ``square-root" form (\ref{sqrt}).
Different definitions of nonabelian Born-Infeld (NBI) lagrangians
are possible, regarding the way of tracing the group indices. The
symmetrized trace version of Tseytlin \cite{t,bre,dg} is often
regarded as the one which describes the non-derivative
approximation of string theory; however, there are other proposals
\cite{oth}. Usually NBI lagrangians cannot be put in the closed
form in the component fields $F_{\mu\nu}^a$.

Following Gal'tsov and Kerner \cite{gk}, in this paper we will
analyze the simplest version of NBI action in which the trace
over the group indices is done under the square-root sign.
Gal'tsov and Kerner found particle-like finite energy solutions
for the NBI action for the $SU(2)$ gauge group. Motivated by this
result and by the fact that the dyonic solutions are of interest
in the brane theory,  we analyze a more general class of
solutions. We also discuss the existence of pure monopole
solutions.

\section{Action and field equations}

The initial point of our analysis is the following nonabelian
Born-Infeld action in Minkowski space:
\be S = {1\over 4\pi}\int  d^4x(1-\car )\ ,\label{NBI}\ee
\noindent where $\car $ is defined as
\be\car =\sqrt{ 1+{1\over 2}F_{\mu\nu}^aF^{\mu\nu a}-
{1\over 16}F_{\mu\nu}^a{F^*}^{\mu\nu a}}\ .\label{R} \ee
\noindent We put that the maximal field
strength equals unity,  $\b=1$. Lorentz inices $\mu $, $\nu $ run
from 0 to 3 and we will often split them into the temporal part 0
and the spatial part, $i$, $j$ = 1, 2, 3. The signature which we
use is $(-,+,+,+)$. $F_{\mu\nu}^a$ are the field strengths of the
$SU(2)$ gauge group,
\be F_{\mu\nu}^a=\partial _\mu A_\nu
^a-\partial _\nu A_\mu ^a+\epsilon ^{abc}A_\mu ^bA_\nu ^c\ ,\ee
\noindent with $a$, $b$ = 1, 2, 3. The equations of motion which
follow from the NBI action (\ref{NBI}) are
\be D_\mu P^{\mu\nu}=0\ ,\label{eqP}\ee
where $P_{\mu\nu}$ are the "displacements"  defined by
\be P_{\mu\nu}^a={\partial\cl
\over\partial F^{\mu\nu a}}={\, F^{\mu\nu a}-G{F^*}^{\mu\nu a}
\over \car}\ \ .\ee
The quantity ${F^*}$ denotes the Hodge-dual of $F$
\be {F^*}^{\mu\nu}={1\over 2}\,\epsilon^{\mu\nu\rho\sigma}
F_{\rho\sigma}\ ,\ee
and we use the shorthand notation
\be G={1\over 4}\, F_{\mu\nu}^a{F^*}^{\mu\nu a}\ .\label{FG}\ee
The equations of motion  (\ref{eqP}) can be
complemented with the Bianchi identities
\be D_\mu {F^*}^{\mu\nu}=0\ .\label{eqF}\ee

It is important to note that NBI theory has the duality symmetry
as BI:
\be F^{\mu\nu}\to {P^*}^{\mu\nu} \ ,\ \ P^{\mu\nu}\to
-{F^*}^{\mu\nu} \label{duality}\ .\ee
Duality invariance can be seen from the vacuum equations
(\ref{eqP}) and (\ref{eqF}), too. It can be used  to generate new
vacuum solutions from the given ones.

The ansatz for the gauge potentials of \cite{gk} was the ``monopole
ansatz",
\be A_0^a=0\ ,\ \ A_i^a=\epsilon \,_{aik}{1-w(r)\over
r}\,{x^k\over r}\label{mon}\ ,\ee
and it describes the purely magnetic configurations. Electric and
magnetic fields are defined by:
\be E_i^a=F_{i0}^a\ ,\ \ B_i^a={1\over 2}\,\epsilon_{ijk}F_{jk}^a\ .\ee

We will generalize the ansatz (\ref{mon}) -- in fact, we will
consider the general spherically symmetric static potential of the
$SU(2)$ group (Witten's ansatz, \cite{vg}). It is given via three
real functions $a_0(r)$, $a_1(r)$ and $w(r)$ of the radial
coordinate $r$. The components of the gauge potential read:
\be A_0^a=a_0(r){x^a\over r}\ ,\ee \be A_i^a=a_1(r){x^a x^i\over
r^2}+\epsilon _{aik}\,{1-w(r)\over r}\,{x^k\over
r}\label{witten}\ .\ee
Here $x^a$, $x^i$ and $x^k$ ($a$, $i$, $k$ = 1, 2, 3)
are the Cartesian coordinates. The field strengths for this
ansatz are
\be \label{E} E_i^a=a_0^\prime \,{x_ix_a\over
r^2}-{a_0w\over r}\,{x_ix_a-\delta _{ia}r^2\over r^2}\ ,\ee
\be\label{B} B_i^a=-2\delta _{ia}\,{1-w\over r^2}+{(1-w)^2\over
r^2}\,{x_ix_a\over r^2}+{\Bigl({1-w\over r^2}\Bigr)}^\prime\,
{x_ix_a-\delta _{ia}r^2\over r}+{a_1w\over
r^2}\,\epsilon_{iak}x_k\ee
and prime denotes the derivative ${d\over dr}$\ . The square root
${\car}$ becomes
\be \car =\sqrt{ 1+{(1-w^2)^2\over r^4}+2{{w^\prime}^2\over r^2}+ 2{
a_1^2w^2\over r^2} -2{a_0^2w^2\over r^2}
-{a_0^\prime}^2-{{[a_0(1-w^2)]^\prime } ^2\over r^4}\label{RR} }\
\ . \ee

In order to find the equations for $a_0(r)$, $a_1(r)$ and $w(r)$
we can consider the condition of extremality of the action or
introduce the ansatz (\ref{witten}) into (\ref{eqP})--(\ref{eqF}).
After the integration of angular variables, the action is
proportional to the lagrangian $L$,
\be L=\int _0^\infty r^2(\car -1)dr\ .\ee
Varying the unknown functions $a_0$, $a_1$ and $w$,
we obtain the set of the equations:
\be w^2a_1=0 \ ,\label{eqa1}\ee
\be (1-w^2){\left( {[a_0(1-w^2)]^\prime \over
r^2\car }\right)}^\prime ={2w^2a_0 \over
\car}-{\left({r^2a_0^\prime\over\car }\right)}^\prime
\label{eqw}\ ,\ee
\be wa_0\,{\left({[a_0(1-w^2)]^\prime \over
r^2\car }\right)}^\prime =-{2w(1-w^2) \over r^2\car }-{\Bigl({
2w^\prime\over\car}\Bigr)}^\prime -{wa_0^2\over\car}+{wa_1^2\over
\car} \label{eqa0}\ .\ee

\section{NBI dyons}

The system of equations (\ref{eqa1}--\ref{eqa0}) is a complicated
nonlinear system. We will search for particular solutions of this
system with finite energy. The energy of the static
configurations is equal to the negative value of the lagrangian,
\be M=\int _0^\infty r^2(1-\car )dr\ .\label{energy}\ee
The convergence of this integral on both boundaries imposes
restrictions on the asymptotic behavior of the functions $a_0$,
$a_1$ and $w$, which we will discuss later.

Let us first consider the simplest equation, (\ref{eqa1}): it
implies that either $w(r)=0$ or $a_1(r)=0$. But one can see rather
easily that the configuration $w(r)=0$, $a_1(r)\neq 0$ is  gauge
equivalent to the configuration $w(r)=0$, $a_1(r)=0$. Indeed, for
$w(r)=0$ we obtain that (\ref{eqa1}) and (\ref{eqa0}) are
identically fulfilled, leaving $a_1(r)$ undetermined. This means
that $a_1(r)$ represents the gauge freedom. The value of $a_1(r)$
does not influence the values of the field strengths in the case
$w(r)=0$, as can be seen from (\ref{E}--\ref{B}). Therefore, we
will always assume that $a_1(r)=0$ and denote $a_0(r) =a(r)$ in
the following, keeping the indexed notation like $a_0$, $a_1$,
$w_0$ etc. for the coefficients in the power series expansions.

The solutions with $a(r)=0$, $w(r)\neq 0$ were discussed by
Gal'tsov and Kerner in detail. In this case, the equations of
motion reduce to
\be \Bigl({w^\prime\over \car}\Bigr)^\prime =
-{w(1-w^2)\over r^2\car }\ \label{gal}\ ,\ee
and the square root $\car$ to the expression
\be \car =\sqrt{1+{(1-w^2)^2\over r^4}+2{{w^\prime}^2\over r^2}}\ \ .\ee
The simplest solution of (\ref{gal}), $w(r)=\pm 1$,
is the pure gauge.  $w(r)=0$ is also a solution, and it has
the form of the Dirac monopole:  this is
embedded $U(1)$  monopole. Its energy is finite:
\be M_e ={\pi^{3/2}\over 3\Gamma (3/4)^2}\approx 1.2360\ .\ee
There is also an infinite discrete set of finite energy solutions
$w_n(r)$, $n\in \mathbb{N}$, the so-called Gal'tsov-Kerner
glueballs. These solutions can be found numerically using the
condition that function $w(r)$ with the allowed asymptotic forms
at $r\to 0$ and $r\to\infty$ is smooth in the intermediate region.
The asymptotic expansions are:
\bea r\to 0&:&\ w(r)=1-b r^2+O(r^4)\ , \ncr r\to\infty
 &:&\ w(r)=\pm 1+{c\over r}+O\left({1\over r^2}\right)\ . \eea
Let us note that the solutions behaving at infinity as $w(r)\to 0$
are excluded, thus leaving only the configurations with no magnetic
charge. Solutions $w_n(r)$ behave as magnetic dipoles and have
energies which tend to the energy $M_e$ of the embedded monopole
as $n\to\infty$ .

\begin{figure}[!ht]
\center
\includegraphics[height=7cm]{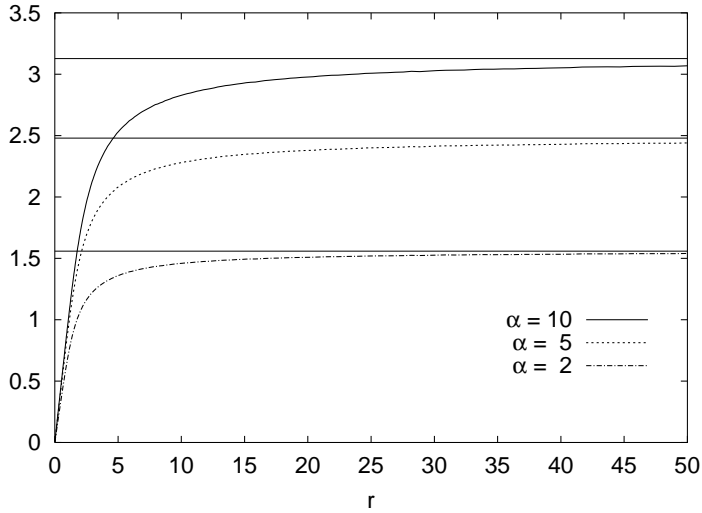}
\caption{Dyon solution for various values of $\a$.} \label{a0}
\end{figure}

The other simple possibility,  $w(r)=0$, $a(r)\neq 0$, is also
nontrivial. The equations of motion in this case reduce to
\be {\Bigl({a^\prime \over r^2\car}\Bigr)}^\prime =
-{\Bigl({r^2a^\prime\over \car}\Bigr)}^\prime\ ,\ee
where now we have
\be \car = \sqrt{ {(1+r^4)(1-{a^\prime }^2)\over r^4}}\ \ .\ee
This equation can be solved explicitly and its solution is a
two-parameter family
\be a(r;\ C, \ \alpha)=C\pm \int_0^r \sqrt{{\a
-1\over \a + r^4}}\,dr\ ,\label{adyon}\ee
where $C$ and $\a$ are the integration constants and $\a >1$.
As the energy and the field strenghts do not depend on $C$ and the
equations are invariant under $a(r)\to -a(r)$, we will take $C=0$
and the + sign in front of the square root. The explicit form of
the solution is given in terms of the elliptic integral \cite{as},
\be a(r;\ \alpha)={1\over 2}(\a -1)^{1/2}\a ^{-1/4}
F\left(\arccos {\sqrt\a -r^2\over \sqrt\a +r^2}\ ,
\ {1 \over 2}\right)\ . \ee
The function $a(r;\ \a )$ is shown in the Figure \ref{a0} for
different values of $\a$. The limiting value of the parameter,
$\a =1$, gives $a(r)=$ const, a configuration
which is gauge equivalent to the embedded monopole $w(r)=0$,
$a(r)=0$. The energy of the solution (\ref{adyon}) is
\be M(\alpha )= {\pi ^{3/2}\over \Gamma (3/4)^2}\, {1\over 2\alpha
^{1/4}}\, \left(1-{\alpha \over 3}\right)\ .\ee
It is unbounded below with the maximum $M_e$ at $\alpha =1$.
We observe that the existence of the electric field decreases the
total energy.

We call the  solution (\ref{adyon}) dyon \cite{jz}, as in the asymptotic
region, $r\to \infty$, the  behavior of the
electric and magnetic fields is given by
\be E_i^a\sim\sqrt{\alpha -1}\,\,{x_ix_a\over r^4}\ ,\ \ B_i^a\sim
-{x_ix_a\over r^4}\ ,\ee
and describes the field strenghts of point-like sources.
The ``electric charge" of the source is proportional to
$\sqrt{\alpha -1}$, while the ``magnetic charge" is 1.

Let us discuss the duals of the forementioned solutions. One
defines the splitting of the displacement tensor in terms of the
vectors $D_i^a$ and $H_i^a$ as
\be P_{i0}^a=D_i^a\ ,\ \ P_{ij}^a=\epsilon _{ijk}H_k^a\ .\ee
The duality transformation (\ref{duality}) can then be reexpressed as
\be E_i^a\to -H_i^a=-{B_i^a-GE_i^a\over \car}\ , \ \  B_i^a\to
D_i^a={E_i^a+GB_i^a\over \car}\ .\ee

In the case of  Gal'tsov-Kerner  glueballs we have $E_i^a =0$,
$G=E_i^aB_i^a=0$, so the duality transforms
\be E_i^a\to -{B_i^a\over\car}\ ,\ \ B_i^a\to {E_i^a\over\car}=0\ .\ee
This means that from the magnetic dipole solution we obtain purely
electric solution, which behaves as a dipole since $\car\to 1$
asymptotically.

In the case of a dyon we see that $G\sim r^{-4}$ and $\car\sim 1$
at infinity. The leading behavior of the transformed
configuration will be
\be B_i^a\sim -{x_ix_a\over r^4}\ ,\ \
E_i^a\sim \sqrt{\a -1}\, \, {x_ix_a\over r^4}\ ,\ee
i. e., the electric and magnetic charges interchange.

\section{General case}

We now turn to the analysis of the general case, $w(r)\neq 0$,
$a(r)\neq 0$. The first condition that we want to impose on our
solutions is finiteness of the energy. This condition restricts
the possible behavior of the functions $w(r)$, $a(r)$ at the
boundaries of the integral (\ref{energy}). If we expand $a(r)$
and $w(r)$ in the power series around $r=0$, we conclude that
(\ref{energy}) converges if there are no poles in the series, i.
e. if they are of the form:
\be a(r)=\sum_0^\infty a_n r^n\ , \ \ 
w(r)=\sum_0^\infty w_n r^n\ . \label{0}\ee
When we analyze the other boundary, $r\to\infty$, we obtain the
similar asymptotics:
\be a(r)=\sum_0^\infty A_n r^{-n}\ ,\ \ w(r)=\sum_0^\infty W_n
r^{-n}\ , \label{inf}\ee
but now the convergence imposes $W_0 A_0=0$.

In order to analyze the relations among the coefficients in
(\ref{0}--\ref{inf}) further, we will assume that the equations of
motion are satisfied order by order in $r$ (or respectively in
$1/r$).

{\bf Case} $r\to 0$. From the equation (\ref{eqw}) we obtain that
$w_0$ must be $\pm 1$ or $0$. As the equations are invariant to
the transformation $w(r)\to -w(r)$ and to $a(r)\to -a(r)$, we
will discuss only $w_0=0$ and $w_0=1$. The similar situation will
repeat in the next case.

For $w_0=0$ the equation (\ref{eqw}) gives: $w_1=w_2=w_3=\dots
=0$; the function $w(r)$ vanishes. At the same time, from
(\ref{eqa1}) we get $a_2=a_3=a_4=0$, $a_5={a_1(a_1^2-1)\over 10}$,
$a_6=a_7=a_8=0$, $a_9= {a_1(a_1^2-1)^2\over 24}$, etc. We also
obtain that $\vert a_1\vert<1$. Thus, both expansions show that
this case corresponds to the dyon solution (\ref{adyon}) with
$a_1=\sqrt{{\a -1\over\a}}\ $.

For $w_0=1$ we get
\bea a(r)&=&a_1 r+a_3 r^3 + O(r^5)\ , \ncr
w(r)&=&1+w_2 r^2+w_4 r^4 +O(r^5)\ , \label{1} \eea
where $$a_3= {8a_1^3w_2+8a_1w_2^3-2a_1w_2\over
10a_1^2-20w_2^2-5}\ ,$$
$$w_4={6w_2^2+a_1^4(2+20w_2^2)+16w_2^4(7+22w_2^2)-
a_1^2(1+42w_2^2+408w_2^4)\over
20(1-a_1^2+4w_2^2)(1-2a_1^2+4w_2^2)}\ ,$$ etc., and
$\vert a_1\vert<{1\over\sqrt{3}}$. We will analyze this asymptotics in
the following, let us just note here that for $a_1=0$ it is the
one obtained in \cite{gk}.

{\bf Case} $r\to\infty$. We consider separately the possibilities
$W_0=0$ and $A_0=0$.

If $W_0=0$, the assumption that the equations of motion are
satisfied order by order in $1/r$ leads to $ W_1=W_2=W_3=\dots
=0$. For coefficients of $a(r)$ we get  $ A_2=A_3=A_4=0$\ ,
$A_5=-{A_1(A_1^2+1)\over 10}$\ , $ A_6=A_7=A_8=0$, $A_9=
{A_1(A_1^2+1)^2\over 24}$, etc. Again, we obtain the
power series expansion of the dyon (\ref{adyon}), in this case
around infinity.

For the second possibility, $A_0=0$, the solutions behave
asymptotically as
\bea a(r)&=&{A_2\over r^2}+{A_3\over r^3}+
{A_4\over r^4}+O\left({1\over r^5}\right) \ncr w(r)&=&1+{W_1\over
r}+{W_2\over r^2}+{W_3\over r^3} +{W_4\over r^4}+O\left({1\over
r^5}\right)\ , \label{2}\eea
where the following relations are fulfilled
$$A_3=A_2W_1\; ,\ \; A_4={18A_2W_1^2-A_2^3\over 20}\; ,\ \;
W_2={6W_1^2-A_2^2\over 8}\ , $$
$$W_3={22W_1^3-9A_2^2W_1\over 40}\; ,\ \;
W_4={17A_2^4-540A_2^2W_1^2+772W_1^4\over 1920}\ \ .$$
\vspace*{1mm}

\begin{figure}[!ht]
\center
\includegraphics[height=4.7cm]{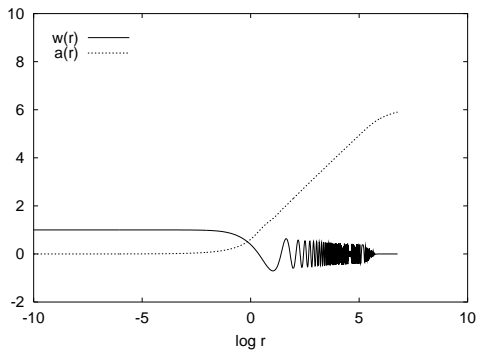}
\caption{Solution for the parameters $w_2=-10$, $a_1=0.5$ and
integration step $h=10^{-3}$. The initial point of integration
is $r_i=10^{-10}$.}
\label{fig2}
\end{figure}

\begin{figure}[!ht]
\center
\includegraphics[height=4.7cm]{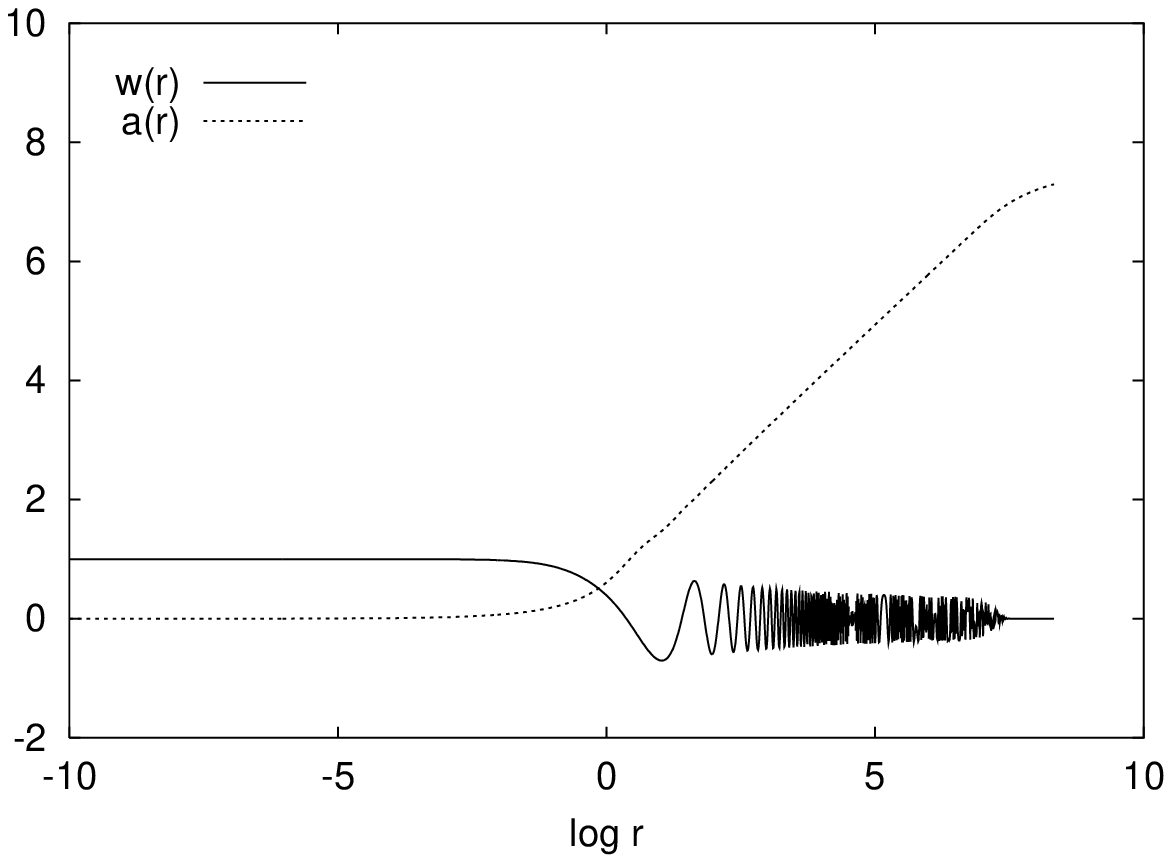}
\includegraphics[height=4.7cm]{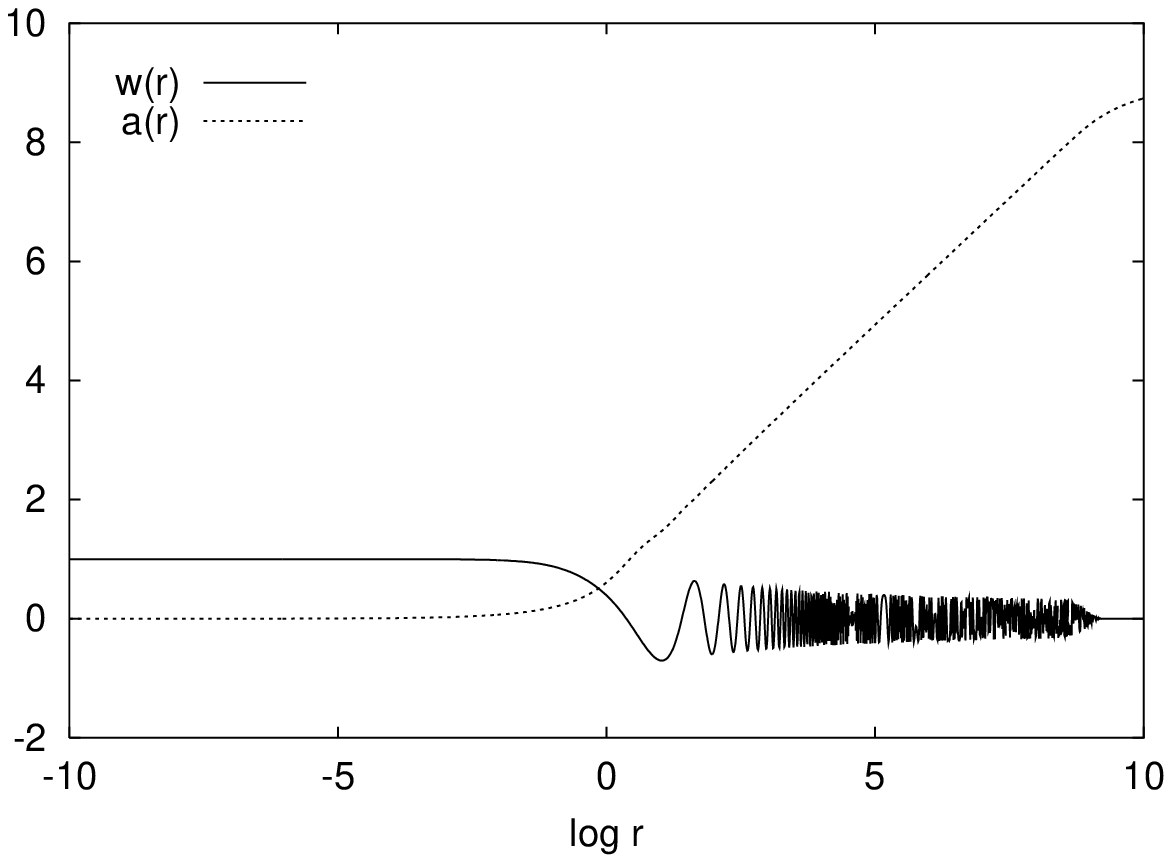}
\caption{Solutions for parameters $w_2=-10$, $a_1=0.5$, $r_i=10^{-10}$
and integration steps $h=10^{-4}$ (left) and $h=10^{-5}$ (right).}
\label{fig3}
\end{figure}

From this analysis we see that, in order to find
new solutions,  we need to join the asymptotics (\ref{1}) and
(\ref{2}) smoothly. Our first attempt was to do the numerical
integration from $r=0$ to the right or from $r=\infty$ to the
left, with the initial conditions defined appropriately. Doing
this, we obtain the generic solution of a typical form shown in
the  Figure \ref{fig2}. The coupling of $a(r)$ and $w(r)$ induces
the oscillations of $w(r)$ which reduce its initial value 1 for
$r=0$ to 0 for $r=\infty$.
This solution is interesting, as it has the behavior of
't Hooft-Polyakov monopole. However, it is numerically unstable: if
we keep the same values of $w_2$, $a_1$ and $r_i$ but decrease the
integration step $h$, we obtain the functions given in the Figure
\ref{fig3}.
The oscillations of $w(r)$ increase to a larger region of $r$,
while the asymptotic value of $a(r)$ changes. We conclude that
the solutions of this type are nonanalytic. Indeed, from the
previous discussion of asymptotics we know that $w(r)\to 0$ as
$r\to \infty$ is compatible  only with $w(r)=0$. Further
numerical analysis of energy  confirms this conclusion: the
values of energy differ for orders of magnitude for different
integration steps and therefore signal that the energy diverges.
We see that in the NBI case, as in the pure Yang-Mills theory,
$w(r)=0$ and $w(r)=1$ are separated by an infinite energy barrier
and it is impossible to find the solution of finite energy which
interpolates between them.

The second possibility for numerical investigation is to start the
integration from both sides $r=0$ and $r=\infty$ with the given
asymptotics, and try to join smoothly the solution in the
intermediate region by varying the parameters $w_2$, $a_1$, $W_1$
and $A_2$. A numerical programme which handles this type of
boundary conditions \cite{nr} was made, and proved to be correct and very
efficient in the simple case of small $a_1$, $A_2$ (glueballs).
However, no new solutions were found using this programme for a
wide range of initial parameters. This might be a consequence of
some weaknesses of the implemented variational procedure
(Newton-Raphson), due to the high dimensionality of the parameter space.
We are, however, inclined to interpret this as a strong numerical
evidence that there are no further finite-energy solutions of the
system (\ref{eqa1}--\ref{eqa0}).

\section{Conclusions}

The set of equations (\ref{eqa1}--\ref{eqa0}), which represent the
equations of motion for the static spherically symmetric
configurations of $SU(2)$ NBI action (\ref{NBI}), is analyzed. The
asymptotic analysis shows that, if one imposes finiteness of
energy, there are only three possible types of solutions:
glueballs, dyons and solutions of the form (\ref{0})--(\ref{1}).

Dyon solutions are of importance in the brane-theory, as they
represent strings ending on three-brane \cite{gib}. The name
dyon, introduced after \cite{jz}, is used in the generalized
sense: there is no Higgs field to determine the unbroken $U(1)$
group. As in the case of Julia-Zee dyon, the electric charge of
this solution is continuous while the magnetic charge is 1.
However, the hope that the components $A_0^a$ of the vector
potential (given via the function $a(r)$) can, through the
nonlinear interaction, take the role of Higgs and counterbalance
the magnetic field to produce the monopole of the 't
Hooft-Polyakov type failed. Instead of the exponential decay,
$a(r)$ induces the oscillations of $w(r)$ with infinite energy.
This could be expected from the fact that the change of the
action from Yang-Mills to NBI does not change the topology of the
fields which are included, necessary for the existence of
monopole \cite{tH}. The solutions of the NBI models with Higgs
fields were discussed in \cite{gpss,bh}.

Finally, let us add that, although the solutions of the third
mentioned type are allowed by the energy considerations, we have a
strong numerical indication that they do not exist. This problem might
deserve further numerical analysis.

{\bf Acknowledgments.}\hspace*{3mm} M. B. wishes to thank
Prof. F. W. Hehl for his hospitality during the stay at the
Institute of Theoretical Physics of the University of Cologne,
when this work was initiated. We also want to thank Profs.
A. Beli\' c and D. Galtsov for the useful comments
concerning the numerical integration. The work was supported in
part by the DAAD grant A/00/17208. The numerics was done at the
IPCF of the Institute of Physics, Belgrade.

\end{document}